\documentstyle[12pt]{article}
\begin{document}
\begin{center}
{\bf{SUPERSYMMETRY IN THERMO FIELD DYNAMICS}}

\vspace{1.0cm}

R.Parthasarathy and R.Sridhar{\footnote{e-mail address:
sarathy,sridhar@imsc.ernet.in}} \\
The Institute of Mathematical Sciences \\
C.P.T.Campus, Taramani Post \\
Chennai 600 113, India. \\
\end{center}

\vspace{0.5cm}

{\noindent{\it{Abstract}}}

By considering the enlarged thermal system including the heat bath, it is shown that this system has supersymmetry
which is not broken at finite temperature. The superalgebra is constructed and the Hamiltonian is
expressed as the anti-commutator of two kinds of super charges. With this Hamiltonian and the thermal vaccum $\mid
0(\beta)>$, this supersymmetry is found to be preserved. 

\vspace{0.5cm}
\newpage 
{\noindent{\bf{I.$\ $ Introduction}}}

\vspace{0.5cm}

Supersymmetry at finite temperature has been studied in detail by Girardello et.al (1981), Van Hove (1982), Das
(1989), Witten (1982), Das and Kaku (1978), Teshima (1983), Umezawa, Matsumoto and Tachiki (1982), and Umezawa
(1993). Nevertheless, the issue of whether supersymmetry is broken or not, at finite temperature has raised some
controversy. Girardello et.al (1981) argued that supersymmetry (SUSY) is broken at positive temperature even when
unbroken at $T = 0$. In response to this, Van Hove (1982), suggested that when a change of an operator under SUSY
transformation at finite temperature is considered, one should take into account the Klein operator. When this
operator is incorporated, Van Hove (1982) shows that the thermal average of this change of an operator is zero for
all $T$, thereby maintaining SUSY at finite temperature. On the other hand, Das (1989) has considered this issue
within the context of the real time formalism or Thermo Field Dynamics (TFD) of Umezawa (Umezawa et.al (1982),
Takahashi and Umezawa (1996)) and concluded that SUSY is broken at finite temperature, by evaluating the statistical
average of the SUSY Hamiltonian at $T\ =\ 0$ as its vacuum expectation value in the 'thermal vacuum $\mid 0(\beta)>$
(where $\beta\ =\ 1/kT$, $k$ being the Boltzmann constant) and showing that it is non-zero at finite temperature.

\vspace{0.5cm}

It is the purpose of this work to examine first the construction of SUSY algebra in TFD and then use it to understand
SUSY at finite temperature by enlarging the Fock space including that of the tilde operators. This procedure is
analogous to the treatment of $\hat{G}$-symmetry by Umezawa and will be explained later. We will exhibit 
mathematical possibilities of preserving supersymmetry at finite temperature in agreement with Van Hove.

\vspace{0.5cm}

{\noindent{\bf{II.$\ $Brief Outline of Thermo Field Dynamics}}}

\vspace{0.5cm}
 
In thermofield dynamics (Umezawa et.al (1982), Umezawa (1993) and Takahashi and Umezawa (1996)), the 'thermal vacuum'
expectation value of an operator is equated to its statistical average. The 'thermal vacuum' is temperature dependent
in a doubled Fock space. This construction procedure leads to the introduction of 'tilde' operators and the doubled
Fock space is a direct product of the two Fock spaces for non-tilde and tilde creation and annihilation operators.
This doubling nature is one of the most fundamental and universal feature of all of the thermal quantum field
formalism. For an ensemble of free Bosons with frequency $\omega$, one has the Hamiltonian
\begin{eqnarray}
H_B&=&\omega a^{\dag}a, \nonumber 
\end{eqnarray}
\begin{eqnarray} 
[a,a^{\dag }]&=&1,
\end{eqnarray}
where $a$ and $a^{\dag}$ are the annihilation and creation operators for Bosons. One introduces the tilde fields
by the Hamiltonian
\begin{eqnarray}
{\tilde{H}}_B&=&\omega {\tilde{a}}^{\dag}{\tilde{a}}, \nonumber 
\end{eqnarray}
\begin{eqnarray}
[\tilde{a},{\tilde{a}}^{\dag}]&=&1, \nonumber
\end{eqnarray}
\begin{eqnarray} 
[a,\tilde{a}]\ =\ [a,{\tilde{a}}^{\dag }]&=&0,
\end{eqnarray}
where $\tilde{a}$ and ${\tilde{a}}^{\dag}$ are the annihilation and creation operators for the tilde Bosonic
fields. The thermal vacuum is given by
\begin{eqnarray}
\mid 0(\beta)>&=&(1-exp(-\beta\omega))^{\frac{1}{2}}exp(e^{-\beta\omega/2}a^{\dag}{\tilde{a}}^{\dag}
)\mid 0,\tilde{0}>,
\end{eqnarray}
where a doubling of Fock space is exhibited. The operators $a,a^{\dag},\tilde{a},{\tilde{a}}^{\dag}$ pertain to
zero temperature. The corresponding operators at finite temperature, namely, $a(\beta), a^{\dag}(\beta),
\tilde{a}(\beta), {\tilde{a}}^{\dag}(\beta)$ are obtained from $a, a^{\dag}, \tilde{a}, {\tilde{a}}^{\dag}$ by
Bogoliubov transformation. It is important to note that while
\begin{eqnarray}
a\mid 0(\beta)>&\neq & 0,
\end{eqnarray}
we have
\begin{eqnarray}
a(\beta)\mid 0(\beta)>&=& 0,
\end{eqnarray}
so that the Fock space at finite temperature is spanned by
\begin{eqnarray}
\mid 0(\beta)>,\ a^{\dag}(\beta)\mid 0(\beta)>,\ {\tilde{a}}^{\dag}(\beta)\mid 0(\beta)>, \ 
\frac{1}{\sqrt{n!}}\ \frac{1}{\sqrt{m!}}\ (a^{\dag}(\beta))^n ({\tilde{a}}^{\dag}(\beta))^m\mid 0
(\beta)>.
\end{eqnarray}  

\vspace{0.5cm}

For an ensemble of Fermions of frequency $\omega$ (say), one has similar relations with the commutator replaced by
anti-commutator, and the Fock space at finite temperature will be spanned by
\begin{eqnarray}
\mid 0(\beta)>,\ f^{\dag}(\beta)\mid 0(\beta)>, \ {\tilde{f}}^{\dag}(\beta)\mid 0(\beta)>, \ f^{\dag}(\beta)
{\tilde{f}}^{\dag}(\beta)\mid 0(\beta)>.  
\end{eqnarray}

A physical interpretation of the doubling of the degrees of freedom , 
namely,$a,a^{\dag}$ and $\tilde{a},{\tilde{a}}^{\dag}
$for Bosons and/or $f, f^{\dag}$ and $\tilde{f}, {\tilde{f}}^{\dag}$ for Fermions is the following. When the vacuum
$\mid 0(\beta)>$ is required to be independent of time, as it should be, we choose the total Hamiltonian for free
Boson fields in TFD as 
\begin{eqnarray}
\hat{H}_B&=& \int d^3k {\omega}_k\{ a^{\dag}_ka_k-{\tilde{a}}^{\dag}_k{\tilde{a}}_k\}.
\end{eqnarray}
To an external stimulus, at $T\neq 0$, certain number of quantum particles are condensed in this system, which is in
thermal equillibrium state with temperature $T$. The absorption of the external energy by the system results in (1)
absorption by excitation of additional quanta and (2) excitation of quantum present in the vacuum, creating a hole.
This is how one has doubling in TFD. In fact, Umezawa, Matsumoto and Tachiki (1982) attribute the excitation of holes
to that in the thermal reservoir. 

In studying the properties of dynamical observables, it is expected to use the non-tilde operators. However, in
studying the symmetry properties of the system, one needs both the tilde and non-tilde operators. This is emphazised
in Umezawa (1993), in studying the spontaneous breakdown of $\hat{G}$ symmetry (that of the Bogoliubov
transformation). In this Reference, the $\hat{G}$ symmetry is defined to be  
spontaneously broken when
$\hat{G}\ \mid 0(\beta)>\ \neq \ 0$ while $[\hat{G},\hat{H}_0]\ =\ 0$. It is to be noted that   
the hat-Hamiltonian is used which has the
tilde operators as well (see sections 7.2.4 and 7.3). 
For a system of free Fermions, the total Hamiltonian (analogue of (8)) is
\begin{eqnarray}
{\hat{H}}_F&=& \int d^3k {\omega}_k\{ f^{\dag}_kf_k-{\tilde{f}}^{\dag}_k{\tilde{f}}_k\}
\end{eqnarray}
with the (Fermion) Fock space in (7).  

\vspace{0.5cm}

{\noindent{\bf{III.$\ $ Supersymmetric Algebra}}}

\vspace{0.5cm}

Following Van Hove (1982), we expect supersymmetry to have non-trivial consequences not only at $T=0$
but also at finite temperature, since all excited states must also some how reflect the supersymmetry
property. In view of this we wish to examine {\it{the possibility of maintaining supersymmetry at finite
temperature as well.}} By considering the enlarged Fock space (6) and (7), and using (8) and (9), we
will arrive at this possibility, agreeing with Van Hove (1982).

We demonstrate this by considering a system of free Bosons and free Fermions. At zero temperature, the
Bosons are described by the creation and annihilation operators
$a^{\dag}$ and $a$ and the corresponding tilde operators  
satisfying the algebra
\begin{eqnarray}
[a,a^{\dag}]=1 &;& [a,a]=0, \nonumber 
\end{eqnarray}
\begin{eqnarray}
[\tilde{a},{\tilde{a}}^{\dag}]=1,\ [\tilde{a},\tilde{a}]=0 &;&
[a,\tilde{a}]=0, \ [a,{\tilde{a}}^{\dag}]=0, 
\end{eqnarray}  
and similarly the Fermions are described by
$f,f^{\dag},\tilde{f},{\tilde{f}}^{\dag}$ satisfying the
algebra
\begin{eqnarray}
\{f,f^{\dag}\}=1 &,& f^2\ =\ (f^{\dag})^2\ =\ 0, \nonumber \\
\{\tilde{f},{\tilde{f}}^{\dag}\}=1 &,& {\tilde{f}}^2\ =\ 
({\tilde{f}}^{\dag})^2\ =\ 0, \nonumber \\
\{f,\tilde{f}\}&=&\{f,{\tilde{f}}^{\dag}\}\ =\ 0.
\end{eqnarray}  

We construct Fermionic(super) charge operators (generators of
Supersymmetry) as
\begin{eqnarray}
Q_+\ =\ af^{\dag} &;& Q_{-}\ =\ a^{\dag}f, \nonumber \\
q_+\ =\ \tilde{a}{\tilde{f}}^{\dag} &;& q_{-} \ =\
{\tilde{a}}^{\dag}\tilde{f}.
\end{eqnarray}
These operators are {\it{nilpotent}}, namely,
$Q^2_{+}=Q^2_{-} =q^2_{+}=q^2_{-}=0$ and convert boson to
fermion and vice-versa when acting on the representative state
$\mid n_B,{\tilde{n}}_B,n_F,{\tilde{n}}_F>$. The elements of
the superalgebra are 
\begin{eqnarray}
Q_{\pm}, q_{\pm}, (N_B+N_F), ({\tilde{N}}_B+{\tilde{N}}_F),
\end{eqnarray}
where $N_B\ =\ a^{\dag}a,\ N_F\ =\ f^{\dag}f,\ {\tilde{N}}_B\
=\ {\tilde{a}}^{\dag}\tilde{a},\ {\tilde{N}}_F\ =\
{\tilde{f}}^{\dag}\tilde{f}$. This algebra is closed
\begin{eqnarray}
\{Q_{+},Q_{-}\}&=& N_B+N_F, \nonumber \\
\{q_{+},q_{-}\}&=&{\tilde{N}}_B+{\tilde{N}}_F, \nonumber \\
\{Q_{+},q_{+}\}\ =\ \{Q_{-},q_{-}\} &=& \{Q_{+},q_{-}\}\ =\ 
\{Q_{-},q_{+}\}\ =\ 0, \nonumber
\end{eqnarray}
\begin{eqnarray} 
[Q_{\pm},(N_B+N_F)]&=&0, \nonumber
\end{eqnarray}
\begin{eqnarray} 
[Q_{\pm},({\tilde{N}}_B+{\tilde{N}}_F]&=& 0, \nonumber
\end{eqnarray}
\begin{eqnarray} 
[q_{\pm},(N_B+N_F)]&=&0, \nonumber 
\end{eqnarray}
\begin{eqnarray} 
[q_{\pm},({\tilde{N}}_B+{\tilde{N}}_F)]&=& 0,
\end{eqnarray}    
satisfying the structure $\{O,O\}=E,\ [O,E]=E,\ [E,E]=E$ for
even (E) and odd (O) operators. The total Hamiltonian for
supersymmetric oscillator $\hat{H}\ =\
a^{\dag}a-{\tilde{a}}^{\dag}\tilde{a}+f^{\dag}f-{\tilde{f}}^
{\dag}\tilde{f}$ is {\it{given by the anti-commutator}}
\begin{eqnarray}
\hat{H}&=&\{Q_{+},Q_{-}\}-\{q_{+},q_{-}\},
\end{eqnarray}
and $Q_{\pm},q_{\pm}$ are Fermionic constants of motion, i.e.,
\begin{eqnarray}
[Q_{\pm},\hat{H}]&=&[q_{\pm},\hat{H}]\ =\ 0.
\end{eqnarray} 
The supersymmetric vacuum state at zero temperature is \\ $\mid
0>\ =\ \mid 0_B,{\tilde{0}}_B,0_F,{\tilde{0}}_F>$ and since
this vacuum is annihilated by $a,\tilde{a},f,\tilde{f}$, it
follows that
\begin{eqnarray}
<0\mid \hat{H}\mid 0>&=&0,
\end{eqnarray}
and
\begin{eqnarray}
Q_{\pm}\mid 0>&=&0, \nonumber \\
q_{\pm}\mid 0>&=&0.
\end{eqnarray}
Thus the supersymmetry constructed in (13) and (14) is not
broken at zero temperature. 

\vspace{0.5cm}

{\noindent{\bf{IV.$\ $ Supersymmetry at finite temperature}}}

\vspace{0.5cm}

At finite temperature, we will exhibit three mathematical
possibilities to examine whether supersymmetry is broken or
not. In view of the structure of vacua at finite temperature
in (6) and (7), we choose the thermal vacuum for the
supersymmetric case as
\begin{eqnarray}
\mid 0(\beta)>&=&\mid
0_B(\beta),{\tilde{0}}_B(\beta),0_F(\beta),{\tilde{0}}_F
(\beta)>.
\end{eqnarray}
The zero-temperature operators $a,\tilde{a},f,\tilde{f}$ are
related to the 
'temperature dependent' operators $a(\beta),
\tilde{a}(\beta), f(\beta),\tilde{f}(\beta)$ which annihilate
the above 'thermal vacuum',
by
the (inverse) Bogoliubov transformation (Takahashi and Umezawa
(1996))
\begin{eqnarray}
a&=& u(\beta)a(\beta)+v(\beta){\tilde{a}}^{\dag}(\beta),
\nonumber \\
\tilde{a}&=&u(\beta)\tilde{a}(\beta)+v(\beta)a^{\dag}(\beta),
\nonumber \\
f&=&U(\beta)f(\beta)+V(\beta){\tilde{f}}^{\dag}(\beta),
\nonumber \\
\tilde{f}&=&U(\beta)\tilde{f}(\beta)-V(\beta)f^{\dag}(\beta),
\end{eqnarray}
where
\begin{eqnarray}
u(\beta)&=& (1-e^{-\beta\omega})^{-\frac{1}{2}}, \nonumber \\
v(\beta)&=& (e^{\beta\omega}-1)^{-\frac{1}{2}}, \nonumber \\
U(\beta)&=&(1+e^{-\beta\omega})^{-\frac{1}{2}}, \nonumber \\
V(\beta)&=&(1+e^{\beta\omega})^{-\frac{1}{2}}.
\end{eqnarray} 

\vspace{0.5cm}

{\noindent{\it{Method.1}}}

\vspace{0.3cm}

In this method, the thermal vacuum is given by (19) and the
Hamiltonian by (15) (involving zero temperature operators).    
Since now we have $a\mid 0(\beta)>\ \neq\ 0,\tilde{a}\mid
0(\beta)>\ \neq\ 0,f\mid 0(\beta)>\ \neq\ 0, \tilde{f}\mid
0(\beta)>\ \neq\ 0$, we need to use the Bogoliubov
transformation (20).   

Then it follows
\begin{eqnarray}
a\mid 0(\beta)>&=&v(\beta){\tilde{a}}^{\dag}(\beta)\mid
0(\beta)>,
\nonumber
\end{eqnarray}
so that,
\begin{eqnarray}
<0(\beta)\mid a^{\dag}a\mid 0(\beta)>&=&v^2(\beta)<0(\beta)
\mid \tilde{a}(\beta){\tilde{a}}^{\dag}(\beta)\mid 0(\beta)>
\nonumber \\
&=&v^2(\beta)<0(\beta)\mid 1+{\tilde{a}}^{\dag}(\beta)\tilde{a}
(\beta)\mid 0(\beta)>\ =\ v^2(\beta), \nonumber 
\end{eqnarray}  
and 
\begin{eqnarray}
\tilde{a}\mid 0(\beta)>&=&v(\beta)a^{\dag}(\beta)\mid 0(\beta)
>, \nonumber 
\end{eqnarray}
so that
\begin{eqnarray}
<0(\beta)\mid {\tilde{a}}^{\dag}\tilde{a}\mid 0(\beta)>&=& v^2
(\beta)<0(\beta)\mid a(\beta)a^{\dag}(\beta)\mid 0(\beta)>
\ =\ v^2(\beta). \nonumber 
\end{eqnarray}
Similarly, we have from
\begin{eqnarray}
f\mid 0(\beta)>&=&V(\beta){\tilde{f}}^{\dag}(\beta)\mid 
0(\beta)>, \nonumber \\
\tilde{f}\mid 0(\beta)>&=& -V(\beta)f^{\dag}(\beta)\mid 
0(\beta)>, \nonumber  
\end{eqnarray}
so that
\begin{eqnarray} 
<0(\beta)\mid f^{\dag}f\mid 0(\beta)>&=& V^2(\beta), \nonumber
\\
<0(\beta)\mid {\tilde{f}}^{\dag}\tilde{f}\mid 0(\beta)>&=&
V^2(\beta). \nonumber
\end{eqnarray}
It then follows from these that
\begin{eqnarray}
<0(\beta)\mid \hat{H}\mid 0(\beta)>&=& <0(\beta)\mid
a^{\dag}a-{\tilde{a}}^{\dag}\tilde{a}+f^{\dag}f-{\tilde{f}}
^{\dag}\tilde{f}\mid 0(\beta)> \nonumber \\ 
& =& 0,
\end{eqnarray}
implying that $\hat{H}$ is invariant under supersymmetry. On
the other hand, 
\begin{eqnarray}
Q_{\pm}\mid 0(\beta)>&\neq & 0, \nonumber \\
q_{\pm}\mid 0(\beta)>&\neq & 0.
\end{eqnarray}
Thus we realize a situation: {\it{while the total Hamiltonian
is supersymmetric invariant, the thermal vacuum is not. We
realize, "spontaneous breakdown of supersymmetry".}} This is
in contrast to the result of Das (1989) in which both his
Hamiltonian and the thermal vacuum are not invariant - a
case of explicit breaking of supersymmetry.  

\vspace{0.5cm}

{\noindent{\it{Method.2}}}

\vspace{0.3cm}

We first motivate the Method.2, by realizing that in the
expressions $<0(\beta)\mid \hat{H}\mid 0(\beta)>$ and those
such as $Q_{\pm}\mid 0(\beta)>$, with $\hat{H}$ given in (15)
and $Q_{\pm},\ q_{\pm}$ in (12), the state vector $\mid
0(\beta)>\ =\ \mid 0_B(\beta),{\tilde{0}}_B(\beta),0_F(\beta),
{\tilde{0}}_F(\beta)>$ is in the doubled Fock space with the
spectrum $\mid n_B(\beta),{\tilde{n}}_B(\beta),n_F(\beta),
{\tilde{n}}_F(\beta)>\ \simeq \ ((a(\beta))^{\dag})^{n_B}\ 
(({\tilde{a}}(\beta))^{\dag})^{{\tilde{n}}_B}\
((f(\beta))^{\dag})^{n_F}\ (({\tilde{f}}(\beta))^{\dag})^{
{\tilde{n}}_F}\mid 0(\beta)>$.  
The states including the vacuum are
temperature dependent. On the other hand, the operators
$\hat{H}$ and $Q_{\pm},q_{\pm}$ are in terms of temperature
independent creation and annihilation operators. In view of
this disparity, it is more appropriate to have operators also
Bogoliubov transformed so that they act on the same Fock
space for examining the symmetry properties of the total
system at finite temperature and this is the reason for
considering Method.2.

First. we consider the Bogoliubov transformed operators
(Takahashi and Umezawa (1996)) which are temperature
dependent, as 
\begin{eqnarray}
a(\beta)&=&u(\beta)a-v(\beta){\tilde{a}}^{\dag}, \nonumber \\
{\tilde{a}}(\beta)&=&u(\beta)\tilde{a}-v(\beta)a^{\dag},
\nonumber \\
f(\beta)&=&U(\beta)f-V(\beta){\tilde{f}}^{\dag}, \nonumber \\
\tilde{f}(\beta)&=&U(\beta)\tilde{f}+V(\beta)f^{\dag},
\end{eqnarray}
which is the inverse of (20) and the functions
$u(\beta),v(\beta),U(\beta),V(\beta)$ are given in (21). It
can be verified that these temperature dependent operators
satisfy the algebra
\begin{eqnarray}
[a(\beta),a^{\dag}(\beta)]&=&1, \nonumber 
\end{eqnarray}
\begin{eqnarray}
[\tilde{a}(\beta),{\tilde{a}}^{\dag}(\beta)]\ =\ 1;\
[a(\beta),\tilde{a}(\beta)]\ =\ [a(\beta), {\tilde{a}}
^{\dag}(\beta)]\ =\ 0, \nonumber 
\end{eqnarray}
\begin{eqnarray}
\{f(\beta),f^{\dag}(\beta)\}=1 &;& f^2(\beta)\ =\
(f^{\dag}(\beta))^2\ =\ 0, \nonumber \\
\{\tilde{f}(\beta),{\tilde{f}}^{\dag}(\beta)\}\ =\ 1 &;&
\{f(\beta),\tilde{f}(\beta)\}\ =\ \{f(\beta),{\tilde{f}}
^{\dag}(\beta)\}\ =\ 0.
\end{eqnarray}
With these, we introduce the temperature dependent super
charges as
\begin{eqnarray}
Q_{+}(\beta)\ =\ a(\beta)f^{\dag}(\beta) &;& Q_{-}(\beta)
\ =\ a^{\dag}(\beta)f(\beta), \nonumber \\
q_{+}(\beta)\ =\ \tilde{a}(\beta){\tilde{f}}^{\dag}(\beta)
&;& q_{-}(\beta)\ =\ {\tilde{a}}^{\dag}(\beta)\tilde{f}
(\beta).
\end{eqnarray}
They are nil-potent. With the number operators,
$(N_B(\beta)+N_F(\beta))\ ;\
(\tilde{N}_B(\beta)+\tilde{N}_F(\beta))$, where
$N_B(\beta)=a^{\dag}(\beta)a(\beta),\
\tilde{N}_B={\tilde{a}}^{\dag}(\beta)\tilde{a}(\beta), \\ 
N_F(\beta)=f^{\dag}(\beta)f(\beta),\ \tilde{N}_F(\beta)=
{\tilde{f}}^{\dag}(\beta)\tilde{f}(\beta)$, they form a closed
super algebra, namely,
\begin{eqnarray}
\{Q_{+}(\beta),Q_{-}(\beta)\}&=&N_B(\beta)+N_F(\beta),
\nonumber \\
\{q_{+}(\beta),q_{-}(\beta)\}&=&\tilde{N}_B(\beta)+\tilde{N}
_F(\beta), \nonumber \\
\{Q_{+}(\beta),q_{+}(\beta)\}&=&\{Q_{+}(\beta),q_{-}(\beta)
\}\ =\ 0, \nonumber \\
\{Q_{-}(\beta),q_{+}(\beta)\}&=&\{Q_{-}(\beta),q_{-}(\beta)
\}\ =\ 0, \nonumber
\end{eqnarray}
\begin{eqnarray} 
[Q_{\pm}(\beta),(N_B(\beta)+N_F(\beta)]&=&0, \nonumber
\end{eqnarray}
\begin{eqnarray}  
[Q_{\pm}(\beta),(\tilde{N}_B(\beta)+\tilde{N}_F(\beta))]&=&0, 
\nonumber
\end{eqnarray}
\begin{eqnarray}  
[q_{\pm}(\beta),(N_B(\beta)+N_F(\beta))]&=&0, \nonumber
\end{eqnarray}
\begin{eqnarray} 
[q_{\pm}(\beta),(\tilde{N}_B(\beta)+\tilde{N}_F(\beta))
]&=&0.
\end{eqnarray}        

Furthermore, we realize the important relation,
\begin{eqnarray}
\{Q_{+}(\beta),Q_{-}(\beta)\}-\{q_{+}(\beta),q_{-}(\beta)\}
=a^{\dag}(\beta)a(\beta)-{\tilde{a}}^{\dag}(\beta)\tilde{a}
(\beta) \nonumber \\
+f^{\dag}(\beta)f(\beta)-{\tilde{f}}^{\dag}(\beta)\tilde{f}
(\beta)\ \equiv \ \hat{H}(\beta).
\end{eqnarray}
This important relation expresses the total Hamiltonian at
finite temperature as anti-commutator of temperature dependent
super charges and therefore $Q_{\pm}(\beta), q_{\pm}(\beta)$
commute with $\hat{H}(\beta)$ showing that they are the
fermionic constants of motion. 

It can be verified upon using (24) that
$\hat{H}(\beta)$ in (28) is the same as $\hat{H}$ in (15)
showing the invariance of the total Hamiltonian under
Bogoliubov transformation. Using, $a(\beta)\mid
0(\beta)>=0\ ;\ \tilde{a}(\beta)\mid 0(\beta)>=0\ ;\\ f(\beta)
\mid 0(\beta)>=0\ ;\ \tilde{f}(\beta)\mid 0(\beta)>=0$, it
follows that
\begin{eqnarray}
<0(\beta)\mid \hat{H}(\beta)\mid 0(\beta)>&=& 0, \nonumber \\
Q_{\pm}(\beta)\mid 0(\beta)>&=&0, \nonumber \\
q_{\pm}(\beta)\mid 0(\beta)>&=&0.
\end{eqnarray} 
Thus, both the total Hamiltonian $\hat{H}(\beta)$ and the
thermal vacuum $\mid 0(\beta)>$ are invariant under
supersymmetry and so supersymmetry is not broken at finite
temperature, in agreement with Van Hove (1982). 

\vspace{0.5cm}

{\noindent{\it{Method.3}}}

\vspace{0.3cm}

The construction of the supersymmetric charges  
$Q_{\pm},q_{\pm}$ in (12) and \\ $Q_{\pm}(\beta),q_{\pm}(\beta)$
in (26) in examining the supersymmetry breaking or not, using
the total Hamiltonian $\hat{H}$ and $\hat{H}(\beta)$ is on
mathematical grounds, in the sense that in thermo field
dynamics, the observables are analysed in terms of non-tilde
operators while the above mathematical procedure includes
tilde operators as well. It is still possible to realize
unbroken supersymmetry at $T\neq 0$ without using the tilde
operators by restricting the super algebra to
\begin{eqnarray}
Q_{\pm}(\beta)&,&  (N_B(\beta)+N_F(\beta)).
\end{eqnarray}
This algebra is closed, namely, 
\begin{eqnarray}
\{Q_{+}(\beta),Q_{-}(\beta)\}&=& N_B(\beta)+N_F(\beta),
\nonumber 
\end{eqnarray}
\begin{eqnarray}
[Q_{\pm}(\beta),(N_B(\beta)+N_F(\beta))]&=&0.
\end{eqnarray}
The Hamiltonian of the system is identified with
\begin{eqnarray}
H(\beta)&=&\{Q_{+}(\beta),Q_{-}(\beta)\}\ =\ a^{\dag}
(\beta)a(\beta)+f^{\dag}(\beta)f(\beta),
\end{eqnarray}
so that
\begin{eqnarray}
[Q_{\pm}(\beta),H(\beta)]&=&0,
\end{eqnarray}
giving the fermionic constants of motion with respect to
$H(\beta)$. The ground state is the thermal vacuum $\mid
0(\beta)>$ as before. Then it follows 
\begin{eqnarray}
<0(\beta)\mid H(\beta)\mid 0(\beta)>&=& 0, \nonumber \\
Q_{\pm}(\beta)\mid 0(\beta)>&=& 0,
\end{eqnarray}
showing that supersymmetry is not broken at finite
temperature. The situation at zero temperature in this case is
obtained by taking the limit $\beta\ \rightarrow \ \infty$. We
have from (20) and (21), 
$u(\beta)\ \rightarrow\ 1\ ;\ v(\beta)\
\rightarrow\ 0\ ;\ a(\beta)\ \rightarrow\ a\ ;\ f(\beta)\
\rightarrow\ f\ ;\ H(\beta)\ \rightarrow\ H\ ;\ \mid
0(\beta)>\ \rightarrow\ \mid 0>$ as $\beta\ \rightarrow\
\infty$ and then we recover the zero temperature case and this
has supersymmetry unbroken.  

\vspace{0.5cm}

{\noindent{\bf{V.$\ $ Summary}}}

\vspace{0.5cm}

We have examined the supersymmetric structure in thermo field
dynamics by considering the enlarged Fock space.
Supersymmetric generators are constructed and three methods,
in which the Hamiltonian is governed by the anti-commutator of
super charges, are studied. Besides realizing spontaneous
breakdown of supersymmetry in Method.1, there are two
possibilities to realize unbroken supersymmetry at finite
temperature. These have well defined zero temperature limit in
which the supersymmetry is not broken. These results are in
agreement with Van Hove. This analysis goes through in the
Lagrangian formulation in TFD and since this is
straightforward, we have not included this. 

\vspace{0.5cm}

{\noindent{\bf{Acknowledgement}}}

\vspace{0.5cm}

Useful discussions with R.Anishetty are acknowledged with
thanks.

\vspace{0.5cm}

{\noindent{\bf{References}}}

\vspace{0.5cm}

\noindent Das A, Physica 1989 {\bf{A158}} 1. \\
Das A and Kaku M, 1978 Phys.Rev. {\bf{D18}} 4540. \\
Girardello L, Grisanuand M T, and Salomonson N P, 1981
Nucl.Phys. {\bf{B178}} 513. \\
Takahashi Y and Umezawa H, 1996 Int.J.Mod.Phys. {\bf{B10}}
1755. \\
Teshima K, 1983 Phys.Lett. {\bf{B123}} 226. \\
Umezawa H, Matsumoto H, and Tachiki M 1982 {\it{Thermo Field
Dynamics}}, North-Holland, Amsterdam. \\
Umezawa H, 1993 {\it{Advanced Field Theory}}, AIP, New York.
\\
Van Hove L, 1982 Nucl.Phys. {\bf{B207}} 15. \\
Witten E, 1982 Nucl.Phys. {\bf{B202}} 253. \\
 
\end{document}